\documentclass[aps,prl,floats,twocolumn]{revtex4}
\usepackage{graphicx}
\begin{document}

\newcommand{\etal} {{\it et al.}}


\title{Electronic texture of the thermoelectric oxide Na$_{0.75}$CoO$_2$}
\author{M.-H. Julien$^{1*}$, C.~de Vaulx$^1$, H.~Mayaffre$^1$, C.~Berthier$^{1,2}$, M.~Horvati\'c$^{2}$, V.~Simonet$^3$, J.~Wooldridge$^4$,
G.~Balakrishnan$^4$, M.R.~Lees$^4$, D.P.~Chen$^5$, C.T.~Lin$^5$, P.~Lejay$^3$}
\affiliation{$^1$Laboratoire de Spectrom\'etrie Physique,
UMR5588 CNRS and Universit\'e J. Fourier - Grenoble, 38402 Saint Martin d'H\`{e}res, France} \affiliation{$^2$Grenoble High Magnetic Field
Laboratory, CNRS, BP166, 38042 Grenoble, France} \affiliation{$^3$Institut N\'eel, CNRS/UJF, BP166, 38042 Grenoble Cedex 9, France}
\affiliation{$^4$Department of Physics, University of Warwick, Coventry CV4 7AL, UK} \affiliation{$^5$Max-Planck-Institut for Solid State
Research, Heisenbergstrasse 1, 70569 Stuttgart, Germany}


\begin{abstract}

From $^{59}$Co and $^{23}$Na NMR, we demonstrate the impact of the Na$^+$ vacancy ordering on the cobalt electronic states in
Na$_{0.75}$CoO$_2$: at long time scales, there is neither a disproportionation into 75\% Co$^{3+}$ and 25\% Co$^{4+}$ states, nor a
mixed-valence metal with a uniform Co$^{3.25+}$ state. Instead, the system adopts an intermediate configuration in which 30~\% of the
lattice sites form an ordered pattern of localized Co$^{3+}$ states. Above 180~K, an anomalous mobility of specific Na$^+$ sites is found
to coexist with this electronic texture, suggesting that the formation of the latter may contribute to stabilizing the Na$^+$ ordering.
Control of the ion doping in these materials thus appears to be crucial for fine-tuning of their thermoelectric properties.

\end{abstract}
\maketitle

Complexity, which underlies many physical properties of correlated electron systems~\cite{complexity}, often results from the spatial
modulation of electronic states. Besides the tendency to form ordered patterns (such as stripes in high temperature superconductors),
electrostatic interactions with dopant ions are increasingly recognized as a source of electronic inhomogeneity~\cite{dopant,CaSrRu}.
Because the sodium ions in Na$_x$CoO$_2$ are mobile and can order, these battery materials are emerging as exceptional candidates for
exploring such phenomena~\cite{Foo}. One of the most important questions regarding this system is whether the intriguing physical
properties of metallic phases at $x\gtrsim0.6$ are related to a coupling between the electronic degrees of freedom and the spatial
distribution of Na$^+$ vacancies~\cite{Foo,TEP2,Zhang05,Meng05,Wang07,Merino06,Marianetti07,Gao07,Roger07,Chou07}. On the one hand, long
range sodium vacancy ordering has been reported but the ordering pattern appears to be controversial for the most studied concentration
$x=0.75$~\cite{Zandbergen04,Geck06,Roger07}. On the other hand, electronic 'textures', which may be defined as both the disproportionation
of electronic (spin, charge and possibly orbital) states and the spatial correlation between these states, are much less experimentally
accessible and, not surprisingly, no electronic pattern has yet been resolved. Nevertheless, the observation of two, or more, magnetically
distinct $^{59}$Co sites at the time scale ($\sim$10$^{-6}$~s) of nuclear magnetic resonance (NMR)
~\cite{Ray99,Itoh00,Ning04,Mukha05,Gavilano06,Mukha07} demonstrates that some electronic disproportionation occurs in metallic
Na$_x$CoO$_2$ with $x\sim0.7$. However, because of the difficulty in accurately determining the Na content $x$, and the sensitivity of Na
order to small variations in $x$ or to the synthesis method, a coherent picture of the electronic states has yet to emerge.

We have obtained reproducible NMR spectra of $^{59}$Co nuclei in high quality single crystals of Na$_{0.75}$CoO$_2$, grown in three
different groups~\cite{Lin,Wooldridge}. Our comprehensive characterization of the crystals will be published separately. For the magnetic
field $H\|c$ ({\it i.e.} $\theta=0^\circ$), a typical spectrum (Fig.~1) shows three distinct sites. These are labeled Co1, Co2 and Co3 in
increasing order of their magnetic hyperfine shift $K= K^{\rm orb} + K^{\rm spin}$. The most interesting differences between these sites
lie in both (i) the value of the on-site orbital contribution $K^{\rm orb}$, which is proportional to the (on-site) Van-Vleck
susceptibility, and (ii) the value of $K^{\rm spin}$ ("Knight shift" in metals), which for a given nuclear site $i$ is given by:
\begin{equation}
K^{\rm spin}_{\alpha\alpha,i}= A_{\alpha\alpha}\frac{\chi^{\rm spin}_{\alpha\alpha}(i)}{g\mu_B} + \sum_j B\frac{\chi^{\rm
spin}_{\alpha\alpha}(j)}{g\mu_B},
\end{equation}
where $A$ is the on-site hyperfine coupling, $\chi^{\rm spin}$ is the local spin susceptibility and $j$ stands for the nearest Co
neighbors to which the nucleus $i$ may be coupled {\it via} a transferred hyperfine interaction $B$. $\alpha=c,ab$ is the direction of the
principal axis of $A$ and $\chi$ tensors, along which $H$ is aligned. Equation~1 makes clear that, if $B\neq 0$ and $\chi^{\rm spin}(i)
\neq \chi^{\rm spin}(j)$, the number of NMR sites with different $K^{\rm spin}(i)$ values may be greater than the number of different
magnetic sites ({\it i.e.} of different $\chi(i)$ values). Thus, the three $^{59}$Co NMR sites do not necessarily correspond to three
distinct electronic densities.

\begin{figure}[t!]
\centerline{\includegraphics[width=7.8cm]{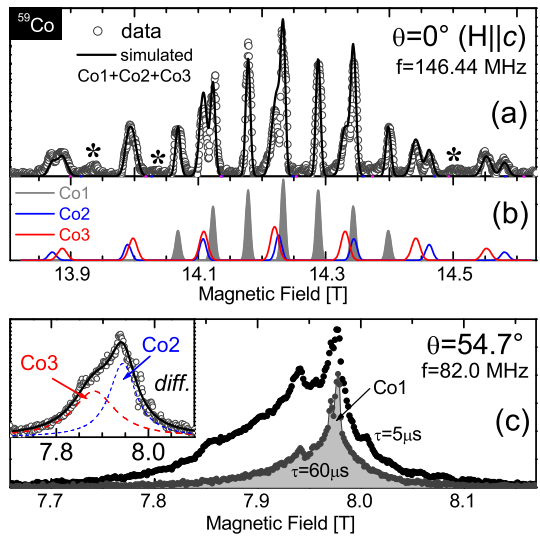}} \caption{(a) Spectrum for $H\|c$ (Circles) and $T=30$~K, with the total simulation
(line) resulting from the Co1, Co2 and Co3 sites displayed in panel (b). Each site consists of seven lines split by the quadrupole
frequency $^{59}\nu_c$. $\nu_c=0.57$~MHz for Co1 is close to the value (0.68~MHz) for Co$^{3+}$ sites in Na$_1$CoO$_2$~\cite{deVaulx05},
and is distinctly lower than $\nu_c=1.22$~MHz for Co2 and 1.14~MHz for Co3. (c) Spectrum for $\theta=54.7^\circ$ where the quadrupole
splitting vanishes, and T=50~K. Co2 and Co3 lines are much broader and more shifted than for $H\|c$. These two lines have a much shorter
spin-spin relaxation time $T_2$ than the Co1 line, so they can be isolated from the latter by subtracting spectra taken at two different
values of the NMR pulse spacing $\tau$ (Inset). Note that, for $\theta=54.7^\circ$, the Co1 line is actually split into Co1a and Co1b
sublines. This effect arises from either in-plane (hyperfine or magnetic) anisotropy or from differences in the Co nearest neighbors at
the Co1 sites. For simplicity, we report here only the properties of the more intense, less shifted, Co1a line (the maximum $K_{\rm iso}$
difference between Co1a and Co1b is 5 to 10 times smaller than the difference between the Co1 and the Co2 or Co3 sites).}
\end{figure}

Additional information on the cobalt sites is provided by the anisotropy and the temperature ($T$) dependence of $K$. $H||ab$ spectra are
too poorly resolved to directly extract the in-plane anisotropy and the $K$ values for all sites. Nevertheless, neglecting the in-plane
anisotropy, $K_{ab}$ can be extracted from a combined knowledge of the line positions for $\theta=54.7^\circ$ and $\theta=0^\circ$, which
define $K_{\rm iso}=\frac{1}{3}(2K_{ab}+K_c)$ and $K_c$ respectively. The values of $K^{\rm orb}$ and of the effective total hyperfine
field ${\cal A^{\rm hf}}$ for each site are then extracted from linear fits of $^{59}K_{ab,c}~vs.~^{23}K$ data, as described
in~\cite{Mukha05}.

For the Co1 site, the values of the orbital shift ($K^{\rm orb}_{ab}$=2.36\%, $K^{\rm orb}_{c}=2.35$\%) are almost the same as those
reported for the Co$^{3+}$ site in Na$_1$CoO$_2$~\cite{deVaulx05,Lang05}. As argued in~\cite{Mukha05,Mukha07}, the isotropic character of
$K_{\rm orb}$ is also consistent with the filled $t_{2g}$ shell of the Co$^{3+}$ ion. Clearly, the Co1 site corresponds to non-magnetic
Co$^{3+}$. This assignment may, at first sight, appear to be in contradiction with $K_{\rm spin}({\rm Co1}) \neq 0$. We cannot rule out
the possibility that the effective valence of the Co1 ion is slightly higher than +3, {\it i.e.} $t_{2g}$ holes have a small but finite
probability of residing on the Co1 sites. However, the quasi-isotropic ${\cal A^{\rm hf}}=28$~kG/$\mu_B$, as well as the different
$T_1~vs.~T$ behavior (see later) for this site would suggest that most of the hyperfine field may be transferred from its magnetic nearest
neighbors ({\it i.e.} the second term of equation~1). Thus, it must be concluded that there are localized sites in the cobalt planes which
are permanently occupied by six $t_{2g}$ electrons.

The Co2 and Co3 sites, on the other hand, are characterized by much larger and anisotropic hyperfine fields values (${\cal A}^{\rm
hf}_{ab}$ = 64 and 154~kG/$\mu_B$ and ${\cal A}^{\rm hf}_{c}$ = 34.5 and 40 kG/$\mu_B$, respectively), showing that the spin density is
much larger. The difference between these sites and Co1 also manifests itself in the much larger $K^{\rm orb}_{ab}$ values (2.55~\% and
2.60~\%, respectively). However, the phenomenological link between $K^{\rm orb}$ anisotropy and the Co valence proposed in
Ref.~\cite{Mukha05} appears not to hold here: $K_{\rm orb}$ values are similar and not strongly anisotropic for Co2 and Co3 ($K^{\rm
orb}_{c}$=2.37 \% and 2.39 \%, respectively). Thus, while the shift difference between the Co2 and Co3 most likely arises from different
electron densities at these sites, we cannot rule out the possibility that part of the difference is due to distinct {\it transferred}
hyperfine fields from their Co nearest neighbors. If present at Co$^{3+}$ sites, the transferred interaction should affect magnetic sites
as well.
%
\begin{figure}[t!]
\centerline{\includegraphics[width=8cm]{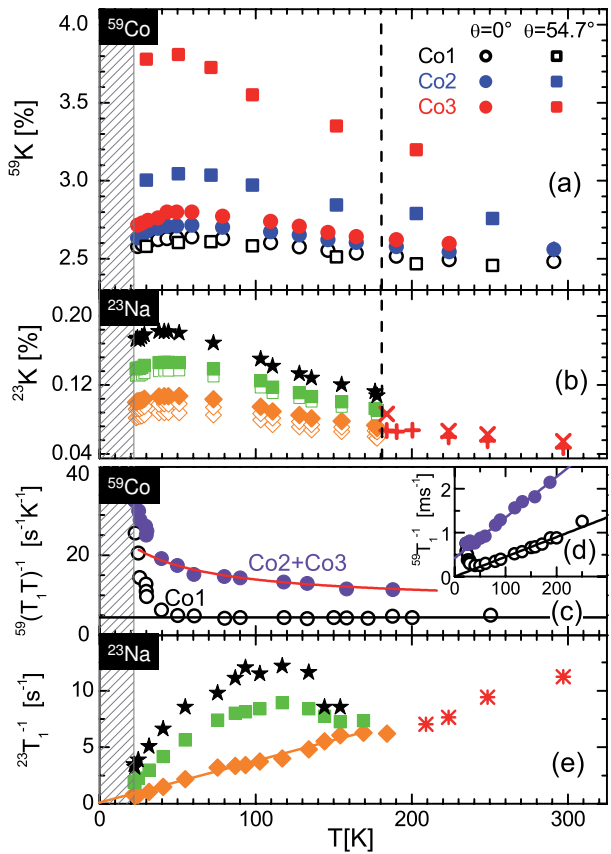}} \caption{(a) $^{59}K$ for the Co1, Co2 and Co3 sites, in two field orientations. The
grey area represents the phase with magnetic order below $T_M=22$~K. (b) $^{23}K$ for $\theta=0^\circ$. The symbol colors refer to lines
in Fig.~3b. The vertical dashed line denotes the temperature of 180~K, where the $^{23}$Na line splits. (c) $(^{59}T_1T)^{-1}$ for the Co1
and the Co2 plus Co3 signals (Co2 and Co3 central lines cannot be separated in $T_1$ measurements performed on central lines with $H\|c$).
Continuous lines are fit results: $(T_1T)^{-1}=4.5$~s$^{-1}$K$^{-1}$ (black), $(T_1T)^{-1}=8+800/(T+35)$~s$^{-1}$K$^{-1}$ (red). Below
$\sim50$~K, critical fluctuations above $T_M$ enhance $T_1^{-1}$, with a progressive loss of Co2 and Co3 signals. (d) Corresponding
$T_1^{-1}$ data. (e) $^{23}T_1^{-1}$ data for the three main $^{23}$Na lines (calibrated from the $^{23}$Na resonance in aqueous NaCl
solution). The symbol colors refer to NMR lines in Fig.~3b. All $T_1$ values were obtained by fitting the recovery curves to standard
expressions for magnetic relaxation.}
\end{figure}

The distinction between magnetic and non magnetic sites is even more striking in the spin-lattice relaxation rates $T_1^{-1}$.
$(T_1T)^{-1}$ is constant at the Co$^{3+}$ (Co1) sites (Fig.~2c). At magnetic sites, on the other hand, $(T_1T)^{-1}$ has a clear $T$
dependence which can be ascribed to a Curie-Weiss law, typical of spin-fluctuation systems. Below $\sim50$~K, critical fluctuations above
$T_M$ enhance $T_1^{-1}$, with a progressive loss of Co2 and Co3 signals.

Although a snapshot of Na$_{0.75}$CoO$_2$ would certainly show 75\% of Co$^{3+}$ and 25\% of Co$^{4+}$ states, this description evidently
does not apply at long time scales: Firstly, the characteristics of the magnetic sites are different from those of the highly oxidized
cobalt (nominally Co$^{4+}$) measured in the $x=0$ member CoO$_2$~\cite{deVaulx07}. Secondly, integration of the NMR signal intensity
(corrected for $T_2$ decay) over the whole spectrum shows that the Co$^{3+}$ sites represent only $30\pm 4$~\% of the total number of
sites throughout the $T$ range. The magnitude of this number, as well as its reproducibility in different samples, demonstrates that
pinning by extrinsic impurities or defects cannot explain the Co$^{3+}$ localization. Therefore, long range magnetic order below 22~K does
not arise from a minority of localized magnetic moments, as often assumed: The spin density is not concentrated on 25\% of the lattice
sites but is spread over $\sim$70\% of them.

\begin{figure}[t!]
\centerline{\includegraphics[width=8cm]{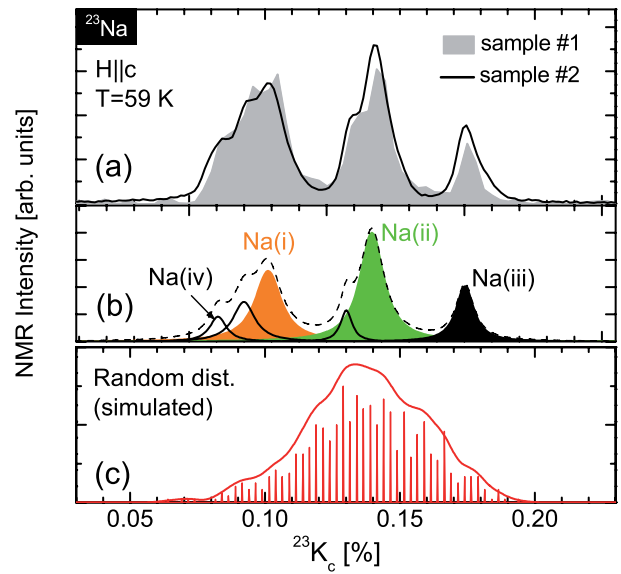}} \caption{(a) Central lines (1/2 $\leftrightarrow$ -1/2 transitions) of the spectrum
in two different single crystals at $T=30$~K. (b) Fit to six lines, having a similar quadrupole frequency $^{23}\nu_c\simeq 1.9$~MHz to
within $\pm$~10~\%. (c) Simulated $^{23}$Na NMR spectrum (line) obtained by a Gaussian broadening of the histogram computed for randomly
distributed magnetic and non-magnetic Co sites. Assuming two different types of magnetic sites further increases the number of lines.
Hyperfine couplings to first neighbors $a$ and to second neighbors $b=a/2$ were assumed to be identical for the Na(1) and Na(2) sites. }
\end{figure}

A knowledge of the local magnetization in the cobalt planes allows us to compute the hyperfine field at the Na sites. Our computer
simulations, an example of which is displayed in Fig.~3c, show that any random distribution of magnetic and non magnetic Co sites produces
a very large number of magnetically distinct $^{23}$Na NMR sites, leading to a broad, featureless, spectrum. This is because the $^{23}$Na
nuclei are coupled to a large number of Co sites: 2 (6) first neighbors and 12 (14) second neighbors for the Na(1) and the Na(2) sites,
respectively. These simulations are at odds with the central transitions of the $^{23}$Na spectrum at $T=59$~K, which consists of only
three clearly separated main resonances, in which six lines with different $K$ values (and negligible $^{23}\nu_c$ differences) are
discernible (four of them are labeled as Na(i,...,iv) on Fig.~3b). The small number of hyperfine fields at the Na sites demonstrates that
the different Co states are spatially ordered~\cite{Mukha04}. At the same time, the specific spectral shape imposes an unprecedented set
of constraints on any model of the Co and Na$^+$ patterns in Na$_{0.75}$CoO$_2$. However, computational uncertainties concerning the 3D
stacking, which is crucial for $^{23}$Na and $^{59}$Co NMR spectra, prevent us from determining whether the in-plane pattern depicted in
Fig.~4~\cite{Roger07}, or another pattern, is correct. The problem should be solved by full DFT calculations of the Na/Co patterns and of
the $^{23}$Na shifts, which are underway.

Surprisingly, the $^{23}$Na spectrum collapses into two very close lines above 180~K (see shifts in Fig.~2b), as if the cobalt planes were
electronically homogeneous. However, the observation of distinct Co1, Co2 and Co3 sites above 180~K tells us that this is not the case.
Thus, the averaging of the hyperfine fields must be due to Na$^+$ ions occupying several distinct sites within the NMR time window. The
Na$^+$ jump frequency has to be larger than the frequency separating the NMR lines with a fine structure ($\sim$10~kHz), but smaller than
the typical frequency of faster probes, such as X-ray diffraction (XRD), for which Na$^+$ order is already well-defined at
$\sim$250~K~\cite{Geck06}. This direct observation of slow Na motion with localization around 180~K rationalizes the anomalies observed in
the $T$ dependence of magnetic~\cite{Schulze07}, lattice~\cite{Geck06}, and possibly charge~(\cite{Ishida07} and references therein),
properties of this material.

Since the NMR timescale is basically identical for $^{59}$Co and $^{23}$Na, the observation that Co electronic differentiation coexists
above 180~K with Na motional averaging is striking. This implies either that electronic disproportionation at the Co sites is completely
unrelated to the Na positions, which would contradict theoretical expectations~\cite{Merino06,Marianetti07,Gao07,Roger07} or that
excursions of the Na$^+$ ions are sufficiently limited so that an electrostatic landscape remains well-defined at long timescales. In this
event, it is likely that the formation of an electronic texture contributes to the final stabilization of the Na$^+$ pattern, as recent
calculations indeed suggest~\cite{Meng05}.
\begin{figure}[t!]
\vspace{-3cm} \centerline{\includegraphics[width=6.5cm]{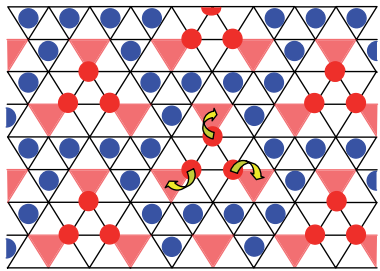}} \caption{Na$^+$ ordering pattern proposed for $x=0.75$ in
Ref.~\cite{Roger07}. Sodium ions at Na(1) positions (red) may hop onto unoccupied neighboring Na(2) positions (red triangles) but sodium
ions at Na(2) positions (blue) cannot because edge-sharing triangles cannot be occupied simultaneously. Our NMR data suggest a similar
selective mobility in Na$_{0.75}$CoO$_2$.}
\end{figure}
Figure 4 illustrates one such possibility where only sodium ions at Na(1) positions may hop onto a neighboring site. Our data suggest that
there is such a contrast in the behavior at the Na sites: There is actually no sign of motion for the $^{23}$Na(i) site since both $K$ and
$1/T_1$ above and below 180~K follow the same trends. $T_1^{-1}$ for this line is approximately linear in $T$ (Fig.~2d), as seen for the
Co$^{3+}$ sites. This similarity strongly suggests that these $^{23}$Na(i) sites are primarily coupled to the Co$^{3+}$ sites. However,
there are 18~\% of $^{23}$Na(i) sites per unit cell (24~\% of the total Na intensity), so there is no simple connection with the 30~\% of
Co$^{3+}$ sites. On the other hand, $^{23}$Na(i)+$^{23}$Na(iv) sites represent 33~\% of the total number of Na$^+$ ions, suggesting that
these sites may correspond to the Na(1) positions~\cite{Roger07,Huang04}. For the $^{23}$Na(ii) and $^{23}$Na(iii) sites,
$^{23}T_1^{-1}~vs.~T$ shows a broad maximum around 100~K (Fig.~2d). This relaxation peak is not seen in the $T_1$ data of any Co site and
is probably caused by anomalous vibrations of the Na$^+$ ions after they have localized on the NMR timescale. We thus anticipate anomalous
phonon modes, possibly analogous to rattling phonons of thermoelectric skutterudites and superconducting pyrochlores~\cite{rattling}.

In conclusion, NMR provides a microscopic view of how Na$^+$ ordering impacts on both the electronic texture and the Na mobility in
Na$_x$CoO$_2$. Since these are essential ingredients of the thermoelectric effect, controlling ion doping in these materials (either by
dilute doping of ions of different mobility or via nanoscale electrochemical manipulations on Na$_x$CoO$_2$ surfaces~\cite{Schneegans07})
should result in an improved thermoelectric performance. More generally, control of ionic textures in battery materials appears to be an
exciting route for tailoring electronic properties.

This work was supported by the ANR NEMSICOM and by the IPMC Grenoble.


\end{document}